\documentclass[letterpaper,12pt]{article}
\usepackage{array}
\usepackage{dcolumn}
\usepackage{float}
\usepackage{natbib}
\usepackage{graphicx}
\usepackage{color}
\usepackage{amsmath}
\usepackage{amsfonts}
\usepackage{amsopn}
\usepackage{amsthm}
\usepackage{amssymb}
\usepackage{lscape}
\usepackage{multirow}

\theoremstyle{definition}

\theoremstyle{remark}

\newtheorem*{rmk.nonumber}{Remark}
\DeclareMathOperator{\pr}{P}
\DeclareMathOperator{\epn}{E}
\DeclareMathOperator{\var}{Var}
\DeclareMathOperator{\plim}{plim}

\DeclareMathOperator{\expit}{expit}

\newcommand{\bmW}{\boldsymbol W}
\newcommand{\bmO}{\boldsymbol O}%
\newcommand{\bmd}{\boldsymbol d}%
\newcommand{\bmb}{\boldsymbol b}%
\newcommand{\bmone}{\boldsymbol 1}%
\newcommand{\bmalpha}{\mbox{\boldmath${\alpha}$}}%
\newcommand{\bmbeta}{\mbox{\boldmath${\beta}$}}%
\newcommand{\bmphi}{\mbox{\boldmath${\phi}$}}%
\newcommand{\bmzero}{\mbox{\boldmath${0}$}}%
\newcommand{\bmnu}{\mbox{\boldmath${\nu}$}}%
\newcommand{\mbI}{\mathbf I}

\renewcommand{\baselinestretch}{1.8}

\begin{document}
\begin{center}
{\LARGE Toward Efficient Estimation of Regional Treatment Effects in Multi-Regional Clinical Trials}
\end{center}
\begin{center}
Zhiwei Zhang$^{1,*}$, Yongwu Shao$^{1}$, Wei Zhang$^{2}$, and Aiyi Liu$^{3}$\\
$^1$Biostatistics, Gilead Sciences, Foster City, California, USA\\
$^2$State Key Laboratory of Mathematical Sciences, Academy of Mathematics and Systems Science, Chinese Academy of Sciences, Beijing, China\\
$^3$Biostatistics and Bioinformatics Branch, Division of Population Health Research, \textit{Eunice Kennedy Shriver} National Institute of Child Health and Human Development, National Institutes of Health, Bethesda, Maryland, USA\\
$^*$zhiwei.zhang6@gilead.com\\
\end{center}

\vspace{0.5cm}
\centerline{\bf Summary}

A multi-regional clinical trial (MRCT) is a single clinical trial conducted in multiple regions simultaneously under a common protocol, which may be used to support parallel submissions to multiple regulatory authorities. For a regional regulatory authority, treatment effects defined specifically for its own region are more relevant to consider than overall treatment effects based on all regions included in an MRCT. A regional treatment effect can be estimated consistently using local data from the region of interest; however, this approach is generally inefficient as it excludes data from other regions and ignores possible similarities between regions. On the other hand, simply pooling data across regions requires strong assumptions and may introduce bias when the required assumptions are not met. Here, we propose a simple and robust approach to estimating a regional treatment effect in a two-arm randomized MRCT. The proposed approach uses a working regression model to incorporate information from baseline covariates as well as data from other regions for improved efficiency. The model accounts for residual regional differences (after adjusting for measured covariates) using interaction terms that describe how the dependence of outcome on treatment and covariates may vary across regions. The adaptive lasso is used to identify null interactions and thus achieve selective borrowing of information from other regions. The resulting regional treatment effect estimator is consistent and asymptotically normal even when the working model is misspecified, and able to improve efficiency over local estimation when there are similarities between regions in the form of null interactions.

\vspace{.5cm}

\section{Introduction}\label{sec:intro}

A multi-regional clinical trial (MRCT) is a single clinical trial conducted in multiple regions simultaneously under a common protocol, with \lq\lq region" defined as a geographical region, a country, or a regulatory region \citep{ich17,fda24}. If properly designed and conducted, an MRCT can be used to support parallel submissions to multiple regulatory authorities, thereby improving the efficiency of global drug development. The primary analysis of an MRCT usually targets an overall treatment effect (to be estimated and tested using pooled data from all regions included in the trial). It is well recognized that treatment effects may vary across regions due to regional differences in patient and disease characteristics, medical practice, and relevant socio-cultural factors. To address such variation, MRCT data analysis is expected to include an evaluation of the \lq\lq consistency" of treatment effects across regions and subpopulations \citep{ich17}. We will refer to this notion of consistency as inter-regional consistency in order to distinguish it from statistical consistency (i.e., asymptotic unbiasedness). Various methods for assessing inter-regional consistency have been proposed and their statistical implications studied \citep[e.g.,][]{c02,pmda07,i10,q10}. At the present time, there seems no clear consensus on how to assess inter-regional consistency and how to interpret and act upon the result of an inter-regional consistency assessment.

This article is concerned with estimation of regional treatment effects (i.e., treatment effects in regional patient populations) using MRCT data. For a regional regulatory authority, treatment effects defined specifically for its own region are arguably more relevant to consider than overall treatment effects. In the context of oncology MRCTs, \citet{fda24} states explicitly that \lq\lq The paramount consideration for FDA when evaluating such oncology trials is whether the results are applicable to the intended use population in the U.S., and to U.S.~standard oncological care." Likewise, Japan's \citet{pmda07} places a great deal of emphasis on treatment effects in the Japanese patient population. If regional treatment effects can be estimated with negligible bias and high precision, such an analysis would be highly relevant to regional regulatory authorities and might help to alleviate concerns about the lack of consensus on how to demonstrate inter-regional consistency.

Due to the aforementioned regional differences, a regional treatment effect of interest may differ from the corresponding overall treatment effect; thus, an overall treatment effect estimator based on pooled data may be (statistically) inconsistent for estimating a regional treatment effect. Regional treatment effects can be estimated consistently using local data (i.e., data from the region of interest); however, this approach is generally inefficient as it excludes data from the other regions and ignores possible similarities between regions. The inefficiency of this commonly used approach may help to explain the relatively low weight of regional treatment effect estimation in the current practice of MRCT data analysis. To improve efficiency over local estimation, \citet{ich17} encourages \lq\lq the search for options for additional pooling of regions based on commonalities, or borrowing information from other regions or pooled regions using an appropriate statistical model". The guideline specifically mentions covariate-adjusted models as a way to account for regional differences and shrinkage estimates as a way to borrow information across regions. Such model-based methods are readily available but may be susceptible to bias when the assumed model is misspecified or when the covariates in the model fail to account for all regional differences. To our knowledge, these issues have not been addressed in the current literature on MRCT data analysis.

In this article, we propose a simple and robust approach to estimating a regional treatment effect in a two-arm randomized MRCT, inspired by recent methodological developments for hybrid control studies \citep{z25,z26}. The proposed approach is consistent under minimal assumptions and able to incorporate information from baseline covariates as well as data from other regions for improved efficiency. Such information is incorporated through a working regression model relating outcome to treatment, region, and other covariates. Correct specification of this model is not required for consistency but may help with efficiency. The model accounts for residual regional differences (after adjusting for measured covariates) using interaction terms that describe how the dependence of outcome on treatment and covariates may vary across regions. Some of these interaction terms may be null (with zero coefficients), representing similarities between regions. Correct identification of null interactions would lead to a reduced model that can be estimated with better precision. Selective borrowing of information from other regions can thus be achieved by applying a variable selection method to the set of interaction terms involving region. Any consistent variable selection method, such as the smoothly clipped absolute deviation method \citep{fl01}, the adaptive lasso \citep{z06} and the minimax concave penalty \citep{z10}, can be used for our purpose. In this work, we focus on the adaptive lasso for simplicity and ease of implementation. It is worth noting that the adaptive lasso remains consistent when the full model is misspecified \citep{l12}; this result is key to the model-robustness of the proposed approach. Under mild regularity conditions, the proposed treatment effect estimator is consistent and asymptotically normal regardless of the (in)correctness of the working model. If there are no null interactions in the model, the proposed estimator is asymptotically equivalent to a covariate-adjusted estimator based on local data. If some interactions are null, the proposed estimator is able to improve efficiency over the local covariate-adjusted estimator, as we demonstrate theoretically and numerically.

The proposed methodology is described in Section \ref{sec:meth}, evaluated by simulation in Section \ref{sec:sim}, and illustrated using real data from an HIV MRCT in Section \ref{sec:app}. Concluding remarks are given in Section \ref{sec:disc}, and technical proofs provided in an appendix. 

\section{Methodology}\label{sec:meth}


Consider a randomized MRCT comparing two treatments in $k+1$ regions, where $k$ is a positive integer. For a generic subject in the trial, let $R\in\{0,1,\dots,k\}$ denote the region in which the patient is enrolled, $\bmW$ a vector of baseline covariates (e.g., patient and disease characteristics measured prior to randomization), $A\in\{0,1\}$ the randomly assigned treatment (1 experimental; 0 control), and $Y$ a fully observed clinical outcome of interest. Randomization implies that $A$ is independent of $(R,\bmW)$ and all other baseline variables. The observed data from $n$ subjects in the trial are denoted by $\bmO_i=(R_i,\bmW_i,A_i,Y_i)$, $i=1,\dots,n$, and conceptualized as independent copies of $\bmO=(R,\bmW,A,Y)$.

Suppose we are interested in estimating the effect of the experimental treatment relative to the control treatment in a specific region, say region 0. Let $\mu_a=\epn\{Y|R=0,A=a\}$, $a\in\{0,1\}$. Commonly used effect measures include the mean difference $\mu_1-\mu_0$, the log mean ratio $\log(\mu_1/\mu_0)$ for outcomes with positive means, and the log odds ratio $\log[\mu_1(1-\mu_0)/\{\mu_0(1-\mu_1)\}]$ for binary outcomes. Each of these can be written as $\delta=g(\mu_1)-g(\mu_0)$, where $g$ is, respectively, the identity function, the log function, or the logit function. In general, $g$ can be any differentiable, strictly increasing function specified by the investigator.

A simple estimator of $\delta$ is readily available from local averages: $\widehat\delta^{\text{Ave-local}}=g(\overline Y_{01})-g(\overline Y_{00})$, where $\overline Y_{0a}=n_{0a}^{-1}\sum_{i=1}^nI(R_i=0,A_i=a)Y_i$, $I(\cdot)$ is the indicator function, and $n_{0a}=\sum_{i=1}^nI(R_i=0,A_i=a)$, $a=0,1$. This estimator is consistent but inefficient because it makes no use of the available data in covariates and from other regions. To overcome these limitations, we follow a g-computation (GC; Robins, 1986) approach based on the identity $\mu_a=\epn\{m_a(\bmW)|R=0\}$, where $m_a(\bmW)=\epn(Y|R=0,A=a,\bmW)$, $a=0,1$. A generic GC estimator of $\mu_a$ is given by $\widehat\mu_a^{\text{GC}}=n_0^{-1}\sum_{i=1}^nI(R_i=0)\widehat m_a(\bmW_i)$, where $n_0=n_{00}+n_{01}$ and $\widehat m_a$ is a generic estimator of $m_a$, $a=0,1$.

To estimate the $m_a$'s, we work with a generalized linear model \citep{mn89} for the conditional distribution of $Y$ given $(R,A,\bmW)$. The model consists of an exponential family (chosen for the nature of $Y$), a canonical link function, and a mean structure given by
\begin{equation}\label{mod.full}
\epn(Y|R,A,\bmW)=h\left((1,A,\bmW',A\bmW')\bmbeta_0+\sum_{r=1}^kI(R=r)(1,A,\bmW',A\bmW')\bmbeta_r\right),
\end{equation}
where $h$ is the inverse link function and $\bmbeta=(\bmbeta_0',\dots,\bmbeta_k')'$ is a vector of unknown parameters. Examples of such models include linear regression for continuous outcomes, logistic regression for binary outcomes, and log-linear Poisson regression for count outcomes. In model \eqref{mod.full}, the $A$-by-$\bmW$ interaction term is optional, but the interactions between $R$ and $(A,\bmW)$ are essential for the model-robustness of the proposed method. Model \eqref{mod.full} implies that $m_a(\bmW)=h((1,a,\bmW',a\bmW')\bmbeta_0)$ and $\mu_a=\epn\{h((1,a,\bmW',a\bmW')\bmbeta_0)|R=0\}$, $a=0,1$. Thus, $\bmbeta_0$ is of primary interest for the purpose of estimating $(\mu_0,\mu_1,\delta)$.

Let $\widehat\bmbeta^{\eqref{mod.full}}$ denote the maximum likelihood estimator (MLE) of $\bmbeta$ obtained by fitting model \eqref{mod.full} to all trial data. Because of the interactions between $R$ and $(A,\bmW)$ in model \eqref{mod.full}, it is easy to see that $\widehat\bmbeta_0^{\eqref{mod.full}}$, the sub-vector of $\widehat\bmbeta^{\eqref{mod.full}}$ corresponding to $\bmbeta_0$, is identical to the MLE of $\bmbeta_0$ that would result from fitting the restricted model
\begin{equation}\label{mod.0}
\epn(Y|R=0,A,\bmW)=h\left((1,A,\bmW',A\bmW')\bmbeta_0\right)
\end{equation}
to the local data in region 0. As such, $\widehat\bmbeta_0^{(1)}$ is based solely on the local data. Substituting $\widehat\bmbeta_0^{\eqref{mod.full}}$ into the GC formula leads to $\widehat\delta^{\text{GC-local}}=g(\widehat\mu_1^{\text{GC-local}})-g(\widehat\mu_0^{\text{GC-local}})$, where
$$
\widehat\mu_a^{\text{GC-local}}=\frac{1}{n_0}\sum_{i=1}^nI(R_i=0)h\left((1,a,\bmW_i',a\bmW_i')\widehat\bmbeta_0^{\eqref{mod.full}}\right),\qquad a=0,1.
$$
This is a standard GC method for covariate adjustment in randomized trials \citep{t08,m09,y23,fda23}. Whether model \eqref{mod.0} is correct or not, $\widehat\delta^{\text{GC-local}}$ is consistent for $\delta$, asymptotically normal, and generally more efficient than $\widehat\delta^{\text{Ave-local}}$ \citep{m09,z19,y23}. On the other hand, because $\widehat\delta^{\text{GC-local}}$ is based on the local data, it is unable to incorporate any relevant information in the available data from the other regions.

One way to incorporate information from other regions is to impose the constraint $\bmbeta_{-0}:=(\bmbeta_1',\dots,\bmbeta_k')'=\bmzero$ in model \eqref{mod.full}, leading to the reduced model
\begin{equation}\label{mod.red}
\epn(Y|R,A,\bmW)=h\left((1,A,\bmW',A\bmW')\bmbeta_0\right).
\end{equation}
Model \eqref{mod.red} implies that $\epn(Y|R,A,\bmW)=\epn(Y|A,\bmW)$. Loosely speaking, this means that $\bmW$ is sufficient to explain any regional differences in the outcome of interest. Let $\widehat\bmbeta_0^{\eqref{mod.red}}$ denote the MLE of $\bmbeta_0$ obtained by fitting model \eqref{mod.red} to all trial data; this can be regarded as a pooled estimator as it does not explicitly adjust for region. The GC estimator of $\delta$ based on $\widehat\bmbeta_0^{\eqref{mod.red}}$ is $\widehat\delta^{\text{GC-pooled}}=g(\widehat\mu_1^{\text{GC-pooled}})-g(\widehat\mu_0^{\text{GC-pooled}})$, where
$$
\widehat\mu_a^{\text{GC-pooled}}=\frac{1}{n_0}\sum_{i=1}^nI(R_i=0)h\left((1,a,\bmW_i',a\bmW_i')\widehat\bmbeta_0^{\eqref{mod.red}}\right),\qquad a=0,1.
$$
This estimator makes effective use of all trial data and is highly efficient if model \eqref{mod.red} is correct. However, it relies on strong assumptions, including the lack of interactions involving region.  When this assumption is violated, $\widehat\delta^{\text{GC-pooled}}$ is generally inconsistent.

Though it may be unrealistic to assume $\bmbeta_{-0}=\bmzero$, it is quite possible that some elements of $\bmbeta_{-0}$ are null (i.e., equal to zero) as some regional differences may disappear after conditioning on $\bmW$. Null elements of $\bmbeta_{-0}$ represent similarities across regions and support information borrowing, while non-null elements of $\bmbeta_{-0}$ require adjustment. It is therefore of interest to identify any null elements of $\bmbeta_{-0}$ and use this information to constrain model \eqref{mod.full} for enhanced estimation precision. This is a general rationale for the variable selection approach to information borrowing, originally proposed for hybrid control studies \citep{z25,z26}. This approach can be implemented using any consistent variable selection method, such as the smoothly clipped absolute deviation method \citep{fl01}, the adaptive lasso \citep{z06}, and the minimax concave penalty \citep{z10}. In this work, we focus on the adaptive lasso, which is relatively simple and readily available in the R package \texttt{glmnet}.

With $m=\dim(\bmbeta_0)$, $\bmbeta_r=(\beta_{r1},\dots,\beta_{rm})'$ and $\widehat\bmbeta_r^{\eqref{mod.full}}=(\widehat\beta_{r1}^{\eqref{mod.full}},\dots,\widehat\beta_{rm}^{\eqref{mod.full}})'$, $r=1,\dots,k$, the adaptive lasso penalty is given by $\lambda\sum_{r=1}^k\sum_{j=1}^m|\beta_{rj}/\widehat\beta_{rj}^{\eqref{mod.full}}|$, where $\lambda$ is a tuning parameter whose value can be chosen through cross-validation. This penalty will be subtracted from the log-likelihood for model \eqref{mod.full}, and the penalized log-likelihood will be maximized with respect to $\bmbeta$ to produce $\widehat\bmbeta^{\text{lasso}}$. Write $\mathcal I=\{(r,j):\beta_{rj}=0,r=1,\dots,k, j=1,\dots,m\}$ and $\widehat{\mathcal I}=\{(r,j):\widehat\beta_{rj}^{\text{lasso}}=0,r=1,\dots,k, j=1,\dots,m\}$. The oracle property of the adaptive lasso \citep{z06} implies that $\pr(\widehat{\mathcal I}=\mathcal I)\to1$ and that $\widehat\bmbeta_0^{\text{lasso}}$ is asymptotically equivalent to the oracle estimator of $\bmbeta_0$, i.e., the restricted MLE under model \eqref{mod.full} with $\{\beta_{rj}:(r,j)\in\mathcal I\}$ fixed at zero. This oracle property continues to hold when model \eqref{mod.full} is mis-specified \citep{l12}, in which case the \lq\lq true" value of $\bmbeta$ is defined as $\bmbeta^*=\plim\widehat\bmbeta^{\eqref{mod.full}}$.

The proposed estimator of $\delta$ is $\widehat\delta^{\text{GC-lasso}}=g(\widehat\mu_1^{\text{GC-lasso}})-g(\widehat\mu_0^{\text{GC-lasso}})$, where
$$
\widehat\mu_a^{\text{GC-lasso}}=\frac{1}{n_0}\sum_{i=1}^nI(R_i=0)h\left((1,a,\bmW_i',a\bmW_i')\widehat\bmbeta_0^{\text{lasso}}\right),\qquad a=0,1.
$$
Whether model \eqref{mod.full} is correct or not, under mild regularity conditions, $\widehat\delta^{\text{GC-lasso}}$ is consistent for $\delta$ and asymptotically normal with (scaled) asymptotic variance
\begin{multline*}
\var\Bigg(\dot g(\mu_1)\left[\frac{I(R=0)\{h((1,1,\bmW',\bmW')\bmbeta_0^*-\mu_1\}}{\pr(R=0)}
+\bmd_1'\bmphi(\bmO)\right]\\
-\dot g(\mu_0)\left[\frac{I(R=0)\{h((1,0,\bmW',0\bmW')\bmbeta_0^*-\mu_0\}}{\pr(R=0)}
+\bmd_0'\bmphi(\bmO)\right]\Bigg),
\end{multline*}
where $\dot g$ (rsp.~$\dot h$) is the derivative function of $g$ (rsp.~$h$), $\bmphi$ is the influence function of $\widehat\bmbeta_0^{\text{lasso}}$, and
$$
\bmd_a=\epn\left\{\left.\dot h\big((1,a,\bmW',a\bmW')\bmbeta_0^*\big)(1,a,\bmW',a\bmW')'\right|R=0\right\},\qquad a=0,1.
$$
A proof of this result is given in Appendix. A plug-in variance estimate can be obtained by replacing the $\var$ operator with sample variance and unknown quantities with empirical estimates. Alternatively, a variance estimate can be obtained using a nonparametric bootstrap procedure, which is computation-intensive but straightforward to implement.

Compared to $\widehat\delta^{\text{GC-pooled}}$, $\widehat\delta^{\text{GC-lasso}}$ is more robust as its consistency requires only regulariy conditions (mainly those required for the oracle property of the adaptive lasso) and no modeling assumptions. Compared to $\widehat\delta^{\text{GC-local}}$, which is also model-robust, $\widehat\delta^{\text{GC-lasso}}$ may have an efficiency advantage, depending on the size of $\mathcal I$ and the correctness of model \eqref{mod.full}. If $\mathcal I$ is empty (i.e., no null interactions involving region), the oracle property of the adaptive lasso implies that $\widehat\bmbeta_0^{\text{lasso}}$ is asymptotically equivalent to $\widehat\bmbeta_0^{\eqref{mod.full}}$ and, therefore, $\widehat\delta^{\text{GC-lasso}}$ is asymptotically equivalent to $\widehat\delta^{\text{GC-local}}$. This makes intuitive sense because methods for information borrowing always require some type of similarities; they are not expected to be helpful when the required similarities are absent. If $\mathcal I$ is non-empty and model \eqref{mod.full} is correct, we demonstrate in Appendix that $\widehat\delta^{\text{GC-lasso}}$ is indeed more efficient than $\widehat\delta^{\text{GC-local}}$. If $\mathcal I$ is non-empty and model \eqref{mod.full} is incorrect, one may still expect an efficiency advantage for $\widehat\delta^{\text{GC-lasso}}$, which effectively incorporates data from other regions. A theoretical proof for this conjecture is not yet available, but this question will be examined numerically in a simulation study.

\section{Simulation}\label{sec:sim}

This section reports a simulation study that evaluates the GC-lasso method in comparison with the Ave-local, GC-local and GC-pooled methods described in Section \ref{sec:meth} as well as a method based on pooled averages, which estimates $\delta$ with $\widehat\delta^{\text{Ave-pooled}}=g(\overline Y_1)-g(\overline Y_0)$, where $\overline Y_a=\{\sum_{i=1}^nI(A_i=a)\}^{-1}\sum_{i=1}^nI(A_i=a)Y_i$, $a=0,1$. In the GC methods, model \eqref{mod.full} is specified as a logistic (for binary $Y$) or linear (for continuous $Y$) regression model, where $h$ is the inverse logit or identity function, respectively. For all methods, analytical standard errors are used to construct 95\% confidence intervals. The target parameter is taken to be the mean or proportion difference $\delta=\mu_1-\mu_0$.

To avoid unnecessary complications, we consider MRCTs conducted in two regions (i.e., $k=1$). There is little loss of generality in doing so, as region 1 (defined by $R=1$) can be the result of combining multiple lower-level regions. We set $\pr(R=0)=\pr(R=1)=0.5$. The covariate vector $\bmW=(W_1,W_2)'$ follows a bivariate normal distribution in each region. Specifically, given $R=r\in\{0,1\}$, $\bmW\sim N_2(\bmnu_r,\mbI)$, where $\bmnu_0=\bmzero$, $\bmnu_1=(0.5,0)'$, and $\mbI$ is the identity matrix. Treatment assignment follows simple 1:1 randomization, so that $\pr(A=1|R,\bmW)=0.5$, independently across subjects. The outcome variable $Y$ may be binary or continuous, and its conditional distribution given $(R,A,\bmW)$ will be described later in four scenarios. We consider $n=400$ or 800. For each sample size and each specified distribution of $(Y|R,A,\bmW)$, $10^4$ sets of trial data are simulated and analyzed using different estimation methods.

In Scenario B0, $Y$ is binary and follows a standard logistic regression model:
\begin{equation}\tag{Scenario B0}
Y=I\left(U<\expit\big\{(1,A,\bmW',A\bmW')\bmalpha_0+R(1,A,\bmW',A\bmW')\bmalpha_1\big\}\right),
\end{equation}
where $\expit(u)=1/\{1+\exp(-u)\}$, $\bmalpha_0=(0,1,-1,1,0.5,0)'$, and $U$ is uniformly distributed on the unit interval and independent of $(R,A,\bmW)$. We consider different choices for $\bmalpha_1$ of the form $(0\bmone_{m-q}',\bmone_q')'$, where $q$ is an integer between 0 and $m=6$ (inclusive) and $\bmone_j$ is a $j$-vector of 1s. Larger values of $q$ represent greater dissimilarity between regions. This mechanism for generating $Y$ is consistent with model \eqref{mod.full} with $h=\text{expit}$, $\bmbeta_0=\bmalpha_0$, $\bmbeta_1=\bmalpha_1$, and $|\mathcal I|=m-q$. It is inconsistent with model \eqref{mod.red} unless $q=0$. Thus, in this scenario, the GC-local and GC-lasso methods are based on correct working models, while the GC-pooled method has a misspecified working model when $q>0$. The true values of $(\mu_0,\mu_1,\delta)$ are approximately $(0.50, 0.69, 0.19)$, regardless of $q$.

\begin{figure}
\centering
\includegraphics[width=1\textwidth]{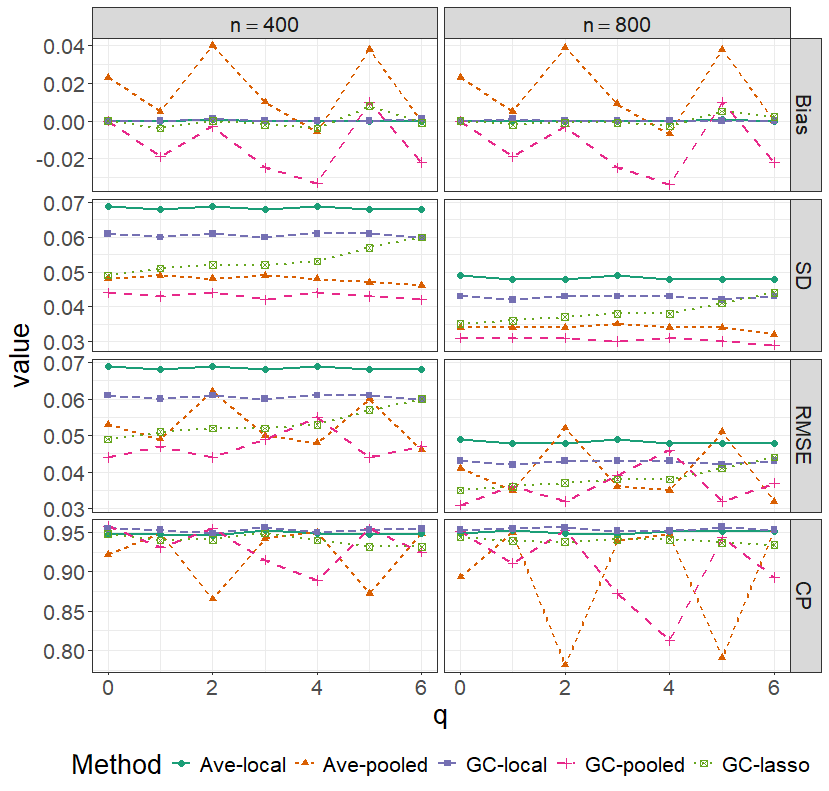}
\caption{Simulation results for Scenario B0 (where $Y$ is binary and model \eqref{mod.full} is correct): empirical bias, standard deviation (SD), root mean squared error (RMSE) and coverage proportion (CP) for estimating $\delta$ using five different estimation methods (see Section \ref{sec:sim} for details).}
\label{sim.rst.B0}
\end{figure}

Figure \ref{sim.rst.B0} shows the simulation results for Scenario B0 in terms of empirical bias, standard deviation, root mean squared error (RMSE), and coverage proportion. As expected, the Ave-local and GC-local methods are (nearly) unbiased while the Ave-pooled and GC-pooled methods can be severely biased, depending on the value of $q$. The GC-lasso method may be slightly biased in some cases (e.g., $q=5$), but the bias is relatively small and appears to diminish with increasing $n$. In terms of variability, the five methods are roughly ordered as follows: $\text{GC-pooled}<\text{Ave-pooled}\le\text{GC-lasso}\le\text{GC-local}<\text{Ave-local}$. Among the three less biased methods, GC-lasso is generally more efficient than GC-local and Ave-local, unless $q=6$ (in which case GC-lasso performs similarly to GC-local, as predicted by asymptotic theory). Asymptotic theory also predicts that GC-lasso performs similarly to GC-pooled when $q=0$; however, a larger sample size than those considered here may be required for this result to take effect. Nonetheless, across different values of $q$, GC-lasso clearly demonstrates a bias advantage over Ave-pooled and GC-pooled as well as an efficiency advantage over Ave-local and GC-local. The RMSE results appear to favor Ave-pooled and GC-pooled, particularly at $n=400$, suggesting that variability dominates bias in this particular scenario. Those two methods exhibit serious under-coverage, while the other three methods have close-to-nominal coverage. The coverage proportion for GC-lasso may be slightly lower than those for Ave-local and GC-local in some cases (e.g., $q=5,6$), but the differences are small and tend to diminish with increasing sample size.

\begin{figure}
\centering
\includegraphics[width=1\textwidth]{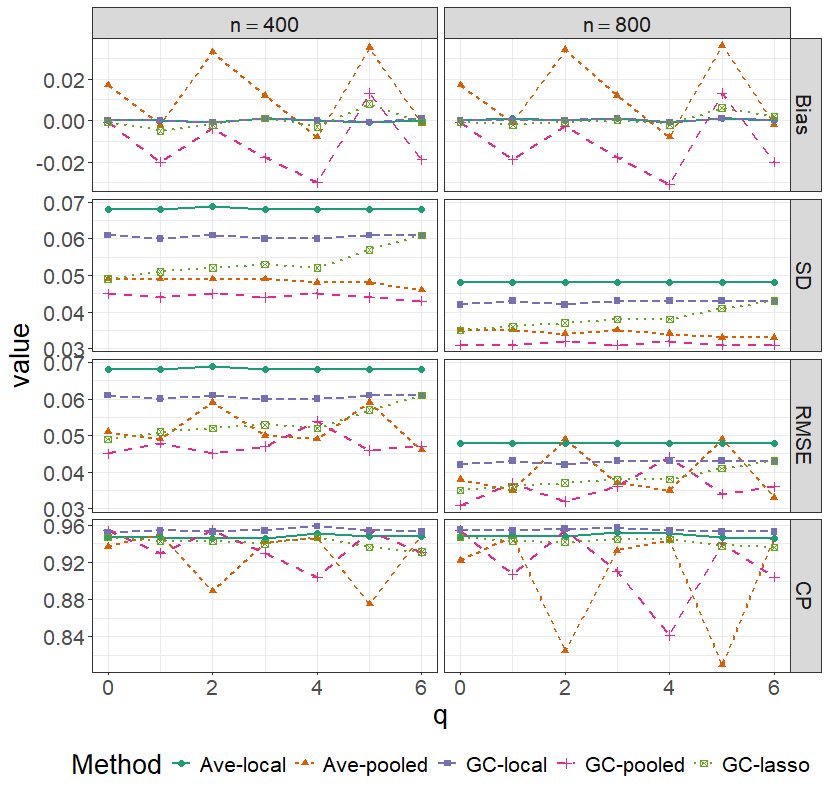}
\caption{Simulation results for Scenario B1 (where $Y$ is binary and model \eqref{mod.full} is incorrect): empirical bias, standard deviation (SD), root mean squared error (RMSE) and coverage proportion (CP) for estimating $\delta$ using five different estimation methods (see Section \ref{sec:sim} for details).}
\label{sim.rst.B1}
\end{figure}

In Scenario B1, $Y$ remains binary and its logistic regression on $(R,A,\bmW)$ contains an interaction between $W_1$ and $W_2$:
\begin{equation}\tag{Scenario B1}
Y=I\left(U<\expit\big\{(1,A,\bmW',A\bmW')\bmalpha_0+R(1,A,\bmW',A\bmW')\bmalpha_2+0.6W_1W_2\big\}\right),
\end{equation}
where $U$ and $\bmalpha_0$ are the same as in Scenario B0 and $\bmalpha_2$ is chosen such that $\bmbeta_1^*=\plim\widehat\bmbeta_1^{\eqref{mod.full}}=\bmalpha_1=(0\bmone_{m-q}',\bmone_q')'$, $q=0,\dots,m$. The values of $\bmalpha_2$ are found numerically by iteratively analyzing larges sets of simulated data with $n=10^5$. Because of the additional interaction between $W_1$ and $W_2$, this data generation mechanism is inconsistent with models \eqref{mod.full} and \eqref{mod.red}. As a result, all three GC methods are now based on incorrect working models. The true values of $(\mu_0,\mu_1,\delta)$ are approximately $(0.52, 0.70, 0.18)$. The simulation results for Scenario B1, shown in Figure \ref{sim.rst.B1}, follow the same patterns as those for Scenario B0.

\begin{figure}
\centering
\includegraphics[width=1\textwidth]{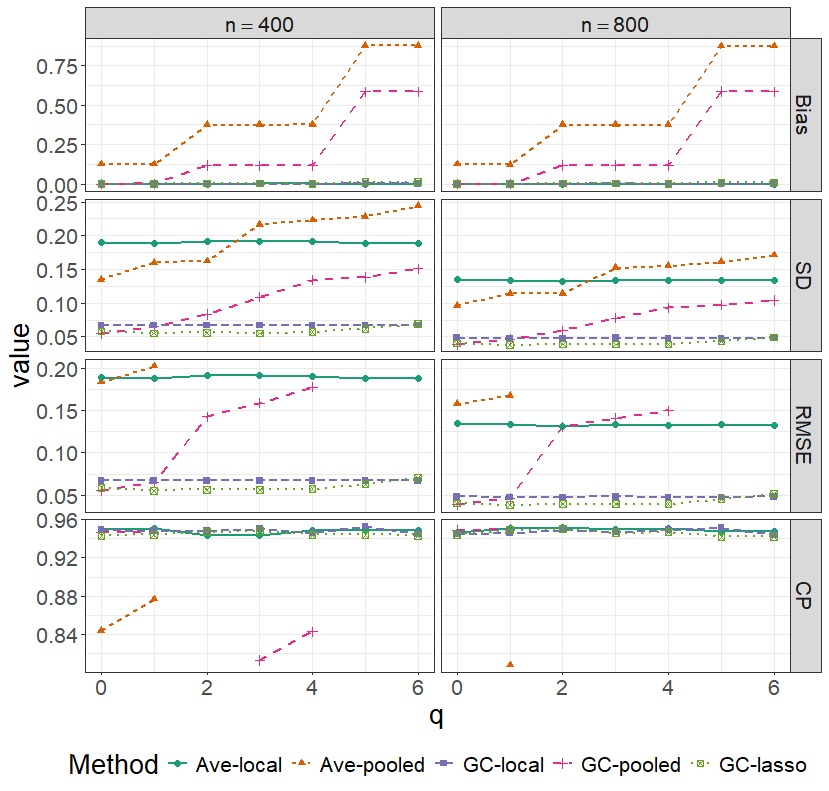}
\caption{Simulation results for Scenario C0 (where $Y$ is continuous and model \eqref{mod.full} is correct): empirical bias, standard deviation (SD), root mean squared error (RMSE) and coverage proportion (CP) for estimating $\delta$ using five different estimation methods (see Section \ref{sec:sim} for details). For readability, some large RMSE values and small CP values are excluded from the plots.}
\label{sim.rst.C0}
\end{figure}

In Scenario C0, $Y$ is continuous and follows a standard linear regression model:
\begin{equation}\tag{Scenario C0}
Y=(1,A,\bmW',A\bmW')\bmalpha_0+R(1,A,\bmW',A\bmW')\bmalpha_1+\varepsilon,
\end{equation}
where $\bmalpha_0$ and $\bmalpha_1$ are the same as in Scenario B0 and $\varepsilon\sim N(0,0.4^2)$, independent of $(R,A,\bmW)$. The value of $\bmalpha_0$ implies that $\mu_0=0$, $\mu_1=1$, and hence $\delta=1$ This data generation mechanism is consistent with model \eqref{mod.full} (with $h$ being the identity function) but inconsistent with model \eqref{mod.red} unless $q=0$. Thus, as in Scenario B0, GC-local and GC-lasso are based on correct models, whereas GC-pooled is based on an incorrect model when $q>0$. Figure \ref{sim.rst.C0} shows the simulation results for Scenario C0, which are qualitatively similar to the previous results with a few notable exceptions concerning Ave-pooled and GC-pooled. Their bias curves appear non-decreasing with $q$ in a stepwise fashion, in contrast to the fluctuations observed in Figures \ref{sim.rst.B0} and \ref{sim.rst.B1}. Their variability also increases with $q$ and generally exceeds the variability of GC-lasso unless $q=0$ (in which case GC-lasso performs similarly to GC-pooled, as predicted by asymptotic theory). It may come as a surprise that the methods based on pooled data (Ave-pooled and GC-pooled) exhibit larger variability than those based on local data (Ave-local and GC-local); however, this can happen when the two regions have large differences, as is apparently the case in this scenario. With larger bias and variability, Ave-pooled and GC-pooled are no longer competitive in terms of RMSE. The RMSE results in Figure \ref{sim.rst.C0} are consistent with theoretical predictions for $q=0$ and $q=m$ and generally in favor of the GC-lasso method.

\begin{figure}
\centering
\includegraphics[width=1\textwidth]{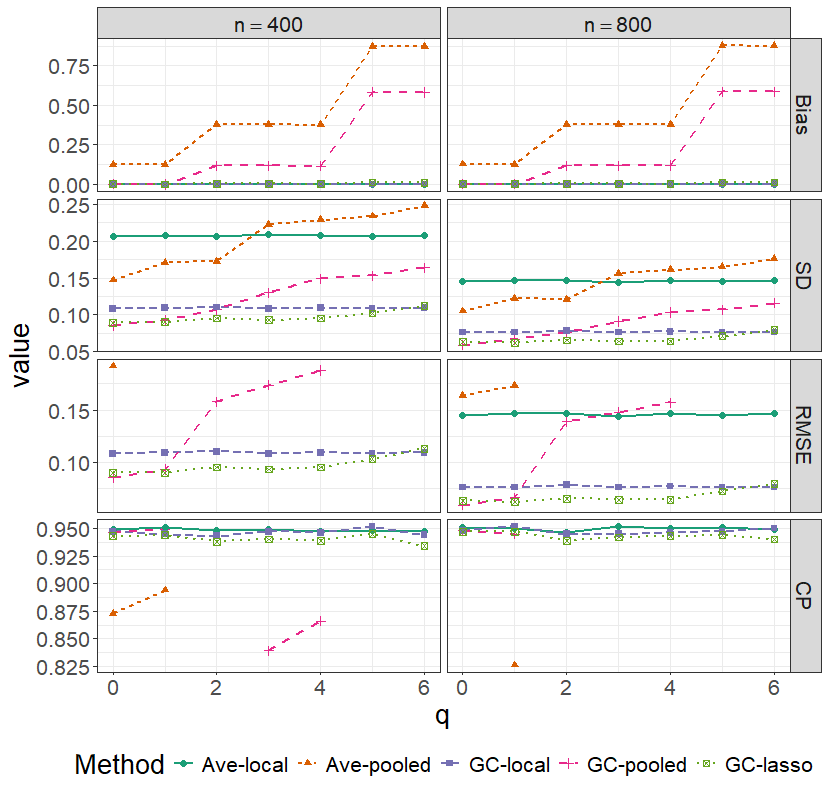}
\caption{Simulation results for Scenario C1 (where $Y$ is continuous and model \eqref{mod.full} is incorrect): empirical bias, standard deviation (SD), root mean squared error (RMSE) and coverage proportion (CP) for estimating $\delta$ using five different estimation methods (see Section \ref{sec:sim} for details). For readability, some large RMSE values and small CP values are excluded from the plots.}
\label{sim.rst.C1}
\end{figure}

The last scenario, C1, is similar to C0 but includes an interaction between $W_1$ and $W_2$:
\begin{equation}\tag{Scenario C1}
Y=(1,A,\bmW',A\bmW')\bmalpha_0+R(1,A,\bmW',A\bmW')\bmalpha_3+0.6W_1W_2+\varepsilon,
\end{equation}
where $\bmalpha_0$ and $\varepsilon$ are the same as in Scenario C0 and $\bmalpha_3$ is chosen such that $\bmbeta_1^*=\plim\widehat\bmbeta_1^{\eqref{mod.full}}=\bmalpha_1=(0\bmone_{m-q}',\bmone_q')'$, $q=0,\dots,m$. Specifically, we set
\begin{multline*}
\bmalpha_3=\bmalpha_1-\left[\epn\{(1,A,\bmW,A\bmW')'(1,A,\bmW,A\bmW')|R=1\}\right]^{-1}\\
\times\epn\left\{0.6W_1W_2(1,A,\bmW,A\bmW')'|R=1\right\}.
\end{multline*}
It is easy to see that $(\mu_0,\mu_1,\delta)$ take the same values as in Scenario C0. Clearly, both model \eqref{mod.full} and \eqref{mod.red} are misspecified; thus, all GC methods are now based on incorrect working models. Figure \ref{sim.rst.C1} shows the simulation results for Scenario C1, which are quite similar to those for Scenario C0.

\section{Illustration}\label{sec:app}

As an illustration, we now apply the methods to a completed MRCT in HIV treatment (NCT02345252; Orkin et al., 2017). This is a randomized, double-blind, placebo-controlled, non-inferiority trial in HIV-1-infected adults who were virally suppressed (HIV-1 RNA $<50$ copies/ml) on emtricitabine, rilpivirine, and tenofovir disoproxil fumarate (FTC/RPV/TDF) for at least 6 months before enrollment and had creatinine clearance of at least 50 ml/min. This trial was conducted to evaluate the efficacy, safety, and tolerability of switching to a fixed-dose combination of emtricitabine, rilpivirine, and tenofovir alafenamide (FTC/RPV/TAF) as compared with remaining on FTC/RPV/TDF. A total of 630 participants were randomized (316 to FTC/RPV/TAF and 314 to FTC/RPV/TDF) and treated at 119 hospitals in 11 countries in North America and Europe. The primary endpoint was the proportion of participants who remained virally suppressed at week 48, with a non-inferiority margin of 8\% (to be applied on the difference scale). Viral suppression was maintained at week 48 for 296 (94\%) participants on FTC/RPV/TAF and 294 (94\%) on FTC/RPV/TDF, with a difference of $-0.3$\% (95\% CI: $-4.2$ to 3.7\%) which easily met the pre-specified non-inferiority criterion.

In this illustration, the regional treatment effect of interest is the average treatment effect in the U.S.~study population as defined by the eligibility criteria of the trial. Most trial participants ($\sim$72\%) were located in the U.S., and the rest in Canada and nine countries in Europe. Based on sample size considerations, the non-U.S.~countries that participated in the trial were combined into one region for the purpose of this illustration. The working model \eqref{mod.full} is specified as a logistic regression model where $\bmW$ represents three baseline covariates: age group ($<40$, 40--54, or $\ge55$), sex, and CD4 cell count. The choice of covariates is based on considerations of prognostic strength and model parsimony (the current model has 20 parameters to estimate). Ten (1.6\%) participants with missing covariate data are excluded from our analysis. Table \ref{hiv.rst} reports the results of our analysis: point estimates and standard errors for estimating $(\mu_0,\mu_1,\delta=\mu_1-\mu_0)$ using the same five methods compared in Section \ref{sec:sim}. The point estimates in Table \ref{hiv.rst} are generally similar across different methods, and the standard errors follow familiar patterns. As expected, the smallest standard errors are produced by the Ave-pooled and GC-pooled methods, which are more susceptible to bias than the other methods. The GC-lasso method produces standard errors smaller or similar to those of the Ave-local and GC-local methods, consistent with theoretical predictions. Taken together, the results in Table \ref{hiv.rst} are consistent with the findings of \citet{o17} and provide further evidence for the efficacy of FTC/RPV/TAF in the U.S.~patient population.

\renewcommand{\baselinestretch}{1.5}
\begin{table}[htbp]
\caption{Analysis of HIV trial data: point estimates (standard errors) of $(\mu_0,\mu_1,\delta=\mu_1-\mu_0)$ from five different estimation methods (see Section \ref{sec:app} for details).}\label{hiv.rst}
\newcolumntype{d}{D{.}{.}{2}}
\begin{center}
\begin{tabular}{lccr}
\hline
\hline
Method&\multicolumn{3}{c}{Viral Suppression at Week 48 (\%)}\\
\cline{2-4}
&\multicolumn{1}{c}{FTC/RPV/TDF}&\multicolumn{1}{c}{FTC/RPV/TAF}&\multicolumn{1}{c}{Difference}\\
\hline
Ave-pooled&93.8 (1.4)&93.6 (1.4)&-0.3 (2.0)\\
GC-pooled&93.8 (1.5)&93.4 (1.5)&-0.3 (2.0)\\
Ave-local&92.9 (1.7)&92.8 (1.7)&-0.1 (2.4)\\
GC-local&92.9 (1.7)&93.2 (1.8)&0.3 (2.5)\\
GC-lasso&93.3 (1.7)&92.8 (1.6)&-0.5 (2.2)\\
\hline
\end{tabular}
\end{center}
\end{table}

\section{Discussion}\label{sec:disc}

The statistical literature on MRCT data analysis places much emphasis on inter-regional consistency, often framed as a hypothesis testing problem concerning homogeneous or similar treatment effects across regions. This emphasis on inter-regional consistency indicates that regional treatment effects are of great interest to regional regulatory authorities and that regional sample sizes are often inadequate to support estimation of regional treatment effects based on local data alone. This apparent gap makes it particularly important to improve the efficiency of regional treatment effect estimation by incorporating relevant information in covariates and from other regions. If regional treatment effects can be estimated efficiently, inter-regional consistency may conceivably become a lesser issue for regional regulatory authorities.

This article may be viewed as a first step toward efficient estimation of regional treatment effects in MRCTs. The GC-lasso method proposed here is able to leverage covariate data as well as regional similarities (in the form of null interactions) for improved efficiency, without introducing an asymptotic bias due to model misspecification or regional differences. The simulation results demonstrate that the GC‑lasso method has a clear efficiency advantage over estimation methods based on local data as well as a bias advantage over methods based on pooled data. The GC-lasso method is remarkably simple and easy to implement, and thus provides a practical and useful tool for MRCT data analysis.

Several open questions remain regarding the GC‑lasso method. While the method is known to be asymptotically more efficient than GC-local when $\mathcal I$ is non-empty and model \eqref{mod.full} correctly specified, an analogous theoretical result has not yet been established under model misspecification. Another important issue concerns relaxation of the condition $|\mathcal I|>0$, which requires some components of $\bmbeta_{-0}^*$ to be exactly 0. In practice, some components of $\bmbeta_{-0}^*$ may be small in magnitude but not identically 0. To better understand the impact of such small values in $\bmbeta_{-0}^*$, it may be helpful to consider a framework in which $\bmbeta_{-0}^*$ depends on the sample size $n$, with some components converging to 0 as $n\to\infty$. Further investigation along these lines may yield new insights that help to understand or improve the theoretical properties and practical performance of the GC-lasso method.

\section*{Acknowledgement}

Aiyi Liu's research  was supported by the intramural research program of the \textit{Eunice Kennedy Shriver} National Institute of Child Health and Human Development.

\section*{Funding}

Wei Zhang's research was supported by the National Key R\&D Program of China [grant number 2022YFA1004800].

\section*{Ethics}

This research was conducted in an ethical manner consistent with Wiley's guidelines on publication ethics.

\section*{Conflict of Interest}

Zhiwei Zhang and Yongwu Shao are employed by Gilead Sciences.

\section*{Data Availability}

The clinical data analyzed in Section \ref{sec:app} is proprietary and not available for sharing.

\section*{Appendix: Proofs}

We assume that model \eqref{mod.full} and its likelihood equation satisfy the conditions in \citet[Chapter 5]{v98} that guarantee the existence and uniqueness of $\bmbeta^*=\plim\widehat\bmbeta^{\eqref{mod.full}}$ and the asymptotic linearity of (restricted) MLEs. We assume that the conditions in \citet[Theorem 1]{l12} hold for model \eqref{mod.full} so that the adaptive lasso, as applied in Section \ref{sec:meth}, possesses the stated oracle property. We assume that, for some $\epsilon>0$, the class of functions $\{h((1,a,\bmW',a\bmW')\bmbeta):\|\bmbeta-\bmbeta^*\|<\epsilon,a=0,1\}$ is Donsker with a square-integrable envelope \citep{vw96}. We write $P_0$ for the true distribution of $\bmO$, $P_n$ for the empirical distribution of $\{\bmO_i,i=1,\dots,n\}$, and $Q_n=\sqrt n(P_n-P_0)$ for the empirical process based on the observed data. These will be used as integration operators; for example, we have
$$
\widehat\mu_a^{\text{GC-lasso}}=\left.P_n\left\{I(R=0)h\left((1,a,\bmW',a\bmW')\widehat\bmbeta_0^{\text{lasso}}\right)\right\}\right/P_nI(R=0),\qquad a=0,1.
$$

We first demonstrate that $\widehat\delta^{\text{GC-lasso}}$ is consistent for $\delta$ without assuming that model \eqref{mod.full} is correct. It suffices to show that each $\widehat\mu_a^{\text{GC-lasso}}$ is consistent for $\mu_a$, that is,
\begin{equation}\tag{A.1}\label{A.1}
\mu_a=\plim\widehat\mu_a^{\text{GC-lasso}}=\epn\left\{\left.h\left((1,a,\bmW',a\bmW')\bmbeta_0^*\right)\right|R=0\right\},\qquad a=0,1.
\end{equation}
To this end, recall that $\bmbeta_0^*=\plim\widehat\bmbeta_0^{\eqref{mod.full}}$, where $\widehat\bmbeta_0^{\eqref{mod.full}}$ satisfies the likelihood equation
$$
0=P_n\left[I(R=0)\left\{Y-h\left((1,A,\bmW',A\bmW')\widehat\bmbeta_0^{\eqref{mod.full}}\right)\right\}(1,A,\bmW',A\bmW')'\right].
$$
It follows that
\begin{equation*}\begin{aligned}
0&=P_0\left[I(R=0)\left\{Y-h\left((1,A,\bmW',A\bmW')\bmbeta_0^*\right)\right\}(1,A,\bmW',A\bmW')'\right]\\
&=\pr(R=0)\epn\left[\left.\left\{Y-h\left((1,A,\bmW',A\bmW')\bmbeta_0^*\right)\right\}(1,A,\bmW',A\bmW')'\right|R=0\right].
\end{aligned}\end{equation*}
Because $\pr(R=0)>0$, the first two components of the above equation imply
\begin{align}
0&=\epn\left\{\left.Y-h\left((1,A,\bmW',A\bmW')\bmbeta_0^*\right)\right|R=0\right\},\tag{A.2}\label{A.2}\\
0&=\epn\left[\left.A\left\{Y-h\left((1,A,\bmW',A\bmW')\bmbeta_0^*\right)\right\}\right|R=0\right].\tag{A.3}\label{A.3}
\end{align}
Subtracting \eqref{A.3} from \eqref{A.2} yields
\begin{equation}\tag{A.4}\label{A.4}
0=\epn\left[\left.(1-A)\left\{Y-h\left((1,A,\bmW',A\bmW')\bmbeta_0^*\right)\right\}\right|R=0\right].
\end{equation}
Equation \eqref{A.3} can be expanded as follows:
\begin{equation*}\begin{aligned}
0&=\pr(A=1|R=0)\left[\epn(Y|A=1,R=0)-\epn\left\{\left.h\left((1,1,\bmW',\bmW')\bmbeta_0^*\right)\right|A=1,R=0\right\}\right]\\
&=\pr(A=1|R=0)\left[\mu_1-\epn\left\{\left.h\left((1,1,\bmW',\bmW')\bmbeta_0^*\right)\right|R=0\right\}\right],
\end{aligned}\end{equation*}
where we use the independence between $A$ and $(R,\bmW)$ in the second step. Because $\pr(A=1|R=0)>0$, this proves \eqref{A.1} with $a=1$. A similar argument can be used to prove \eqref{A.1} with $a=0$ using equation \eqref{A.4}.

The asymptotic linearity of $\widehat\delta^{\text{GC-lasso}}$ follows from the same property of each $\widehat\mu_a^{\text{GC-lasso}}$. For each $a\in\{0,1\}$, we write
\begin{equation}\tag{A.5}\label{A.5}\begin{aligned}
&\quad\sqrt n(\widehat\mu_a^{\text{GC-lasso}}-\mu_a)\\
&=\sqrt n\left[\frac{P_n\{I(R=0)h((1,a,\bmW',a\bmW')\widehat\bmbeta_0^{\text{lasso}})\}}{P_nI(R=0)}
-\frac{P_0\{I(R=0)h((1,a,\bmW',a\bmW')\bmbeta_0^*)\}}{P_0I(R=0)}\right]\\
&=\sqrt n\left[\frac{P_n\{I(R=0)h((1,a,\bmW',a\bmW')\widehat\bmbeta_0^{\text{lasso}})\}}{P_nI(R=0)}
-\frac{P_n\{I(R=0)h((1,a,\bmW',a\bmW')\widehat\bmbeta_0^{\text{lasso}})\}}{P_0I(R=0)}\right]\\
&\quad+\sqrt n\left[\frac{P_n\{I(R=0)h((1,a,\bmW',a\bmW')\widehat\bmbeta_0^{\text{lasso}})\}}{P_0I(R=0)}
-\frac{P_0\{I(R=0)h((1,a,\bmW',a\bmW')\widehat\bmbeta_0^{\text{lasso}})\}}{P_0I(R=0)}\right]\\
&\quad+\sqrt n\left[\frac{P_0\{I(R=0)h((1,a,\bmW',a\bmW')\widehat\bmbeta_0^{\text{lasso}})\}}{P_0I(R=0)}
-\frac{P_0\{I(R=0)h((1,a,\bmW',a\bmW')\bmbeta_0^*)\}}{P_0I(R=0)}\right]\\
&=:T_1+T_2+T_3
\end{aligned}\end{equation}
and analyze the three terms separately. Firstly,
\begin{equation}\tag{A.6}\label{A.6}\begin{aligned}
T_1&=\frac{-P_n\{I(R=0)h((1,a,\bmW',a\bmW')\widehat\bmbeta_0^{\text{lasso}})\}Q_nI(R=0)}{P_nI(R=0)P_0I(R=0)}\\
&=\frac{-P_0\{I(R=0)h((1,a,\bmW',a\bmW')\bmbeta_0^*)\}Q_nI(R=0)}{P_0I(R=0)P_0I(R=0)}+o_p(1)\\
&=\frac{-\epn\{h((1,a,\bmW',a\bmW')\bmbeta_0^*)|R=0\}Q_nI(R=0)}{P_0I(R=0)}+o_p(1)\\
&=\frac{-\mu_aQ_nI(R=0)}{P_0I(R=0)}+o_p(1).
\end{aligned}\end{equation}
Secondly, by Lemma 19.24 of \citet{v98},
\begin{equation}\tag{A.7}\label{A.7}\begin{aligned}
T_2&=\frac{Q_n\{I(R=0)h((1,a,\bmW',a\bmW')\widehat\bmbeta_0^{\text{lasso}})\}}{P_0I(R=0)}\\
&=\frac{Q_n\{I(R=0)h((1,a,\bmW',a\bmW')\bmbeta_0^*)\}}{P_0I(R=0)}+o_p(1).
\end{aligned}\end{equation}
Lastly, by the delta method,
\begin{equation}\tag{A.8}\label{A.8}
T_3=\bmd_a'\bmphi(\bmO)+o_p(1),
\end{equation}
where $\bmphi$ is the influence function of $\widehat\bmbeta_0^{\text{lasso}}$ and $\bmd_a$ is defined in Section \ref{sec:meth}. Substituting \eqref{A.6}--\eqref{A.8} into \eqref{A.5} leads to
\begin{equation*}
\sqrt n(\widehat\mu_a^{\text{GC-lasso}}-\mu_a)=Q_n\left[\frac{I(R=0)\{h((1,a,\bmW',a\bmW')\bmbeta_0^*)-\mu_a\}}{P_0I(R=0)}+\bmd_a'\bmphi(\bmO)\right]+o_p(1).
\end{equation*}
Applying the delta method once again, we obtain
$$
\sqrt n(\widehat\delta^{\text{GC-lasso}}-\delta)=Q_n\psi(\bmO)+o_p(1),
$$
where
\begin{multline*}
\psi(\bmO)=\dot g(\mu_1)\left[\frac{I(R=0)\{h((1,1,\bmW',\bmW')\bmbeta_0^*-\mu_1\}}{P_0I(R=0)}
+\bmd_1'\bmphi(\bmO)\right]\\
-\dot g(\mu_0)\left[\frac{I(R=0)\{h((1,0,\bmW',0\bmW')\bmbeta_0^*-\mu_0\}}{P_0I(R=0)}
+\bmd_0'\bmphi(\bmO)\right].
\end{multline*}
Therefore, $\widehat\delta^{\text{GC-lasso}}$ is asymptotically normal with asymptotic variance $\var\{\psi(\bmO)\}$.

In what follows, we assume that $\mathcal I$ is non-empty and that model \eqref{mod.full} is correct, and demonstrate the efficiency advantage of $\widehat\delta^{\text{GC-lasso}}$ over $\widehat\delta^{\text{GC-local}}$. Recall that $\widehat\bmbeta_0^{\text{lasso}}$ is asymptotically equivalent to the oracle estimator of $\bmbeta_0$, the restricted MLE under model \eqref{mod.full} with $\{\beta_{rj}:(r,j)\in\mathcal I\}$ fixed at zero. Because model \eqref{mod.full} is assumed correct, $\bmphi(\bmO)$ is the efficient influence function for estimating $\bmbeta_0$ under the oracle model and, therefore, belongs to the tangent space of the oracle model \citep[Theorem 4.3]{t06}. In particular, $\bmphi(\bmO)$ takes the form $\{Y-\epn(Y|R,A,\bmW)\}\bmb(R,A,\bmW)$ for some vector-valued function $\bmb$ of the same dimension as $\bmbeta_0$. As such, $\bmphi(\bmO)$ is uncorrelated with any function of $(R,A,\bmW)$. This observation allows the asymptotic variance of $\widehat\delta^{\text{GC-lasso}}$ to be simplified as follows:
$$
\var\{\psi(\bmO)\}=\var\{s(R,\bmW)\}+\{\dot g(\mu_1)\bmd_1-\dot g(\mu_0)\bmd_0\}'\var\{\bmphi(\bmO)\}\{\dot g(\mu_1)\bmd_1-\dot g(\mu_0)\bmd_0\},
$$
where
$$
s(R,\bmW)=\frac{I(R=0)\left[\dot g(\mu_1)\{h((1,1,\bmW',\bmW')\bmbeta_0^*-\mu_1\}-\dot g(\mu_0)\{h((1,0,\bmW',0\bmW')\bmbeta_0^*-\mu_0\}\right]}{P_0I(R=0)}.
$$
Similarly, the asymptotic variance of $\widehat\delta^{\text{GC-local}}$ can be written as
$$
\var\{s(R,\bmW)\}+\{\dot g(\mu_1)\bmd_1-\dot g(\mu_0)\bmd_0\}'\var\{\bmphi^{\eqref{mod.full}}(\bmO)\}\{\dot g(\mu_1)\bmd_1-\dot g(\mu_0)\bmd_0\},
$$
where $\bmphi^{\eqref{mod.full}}$ is the influence function of $\widehat\bmbeta_0^{\eqref{mod.full}}$. To demonstrate that $\widehat\delta^{\text{GC-lasso}}$ is more efficient than $\widehat\delta^{\text{GC-local}}$, it suffices to show that $\var\{\bmphi(\bmO)\}\le\var\{\bmphi^{\eqref{mod.full}}(\bmO)\}$ in the sense of nonnegative-definiteness.The latter follows from the fact that the oracle estimator of $\bmbeta_0$, which has $\bmphi$ as its influence function, is efficient under the oracle model, whereas $\widehat\bmbeta_0^{\eqref{mod.full}}$ is not efficient when $\mathcal I$ is non-empty.

\end{document}